\documentclass[aps,prl,reprint,superscriptaddress,amsmath,amssymb,floatfix]{revtex4-2}

\usepackage{graphicx}
\usepackage{dcolumn}
\usepackage{bm}
\usepackage[normalem]{ulem}

\usepackage{xcolor}
\usepackage{xr-hyper}
\usepackage{hyperref}
\usepackage{cleveref}
\crefname{figure}{Fig.}{Figs.}
\Crefname{figure}{Fig.}{Figs.}
\makeatletter
\renewcommand{\section}{
  \@startsection
    {section}
    {1}
    {\z@}
    {0.8cm \@plus 1ex \@minus .2ex}
    {-.3333em}
    {\normalfont\normalsize\itshape}
}

\def\@runin@to@section#1#2#3{#1#2#3.\textemdash}
\def\@runin@tos@section#1#2{#1#2.\textemdash}

\makeatother

\begin{document}

\preprint{APS/123-QED}

\title{Robust Strange Metallicity across Attractive and Repulsive Hubbard Models}

\author{Xiaoyue Ma}
 \email{phoenixm@stanford.edu}
\affiliation{Department of Materials Science and Engineering, Stanford University, Stanford, CA 94305, USA
}%
\affiliation{Stanford Institute for Materials and Energy Sciences, SLAC National Accelerator Laboratory and Stanford University, Menlo Park, CA 94025, USA}

\author{Emily Z. Zhang}
 \email{emilyzh@stanford.edu}
\affiliation{Department of Materials Science and Engineering, Stanford University, Stanford, CA 94305, USA
}%
\affiliation{Geballe Laboratory for Advanced Materials, Stanford, CA 94305, USA}

\author{Thomas P. Devereaux}
 \email{tpd@stanford.edu}
\affiliation{Department of Materials Science and Engineering, Stanford University, Stanford, CA 94305, USA
}
\affiliation{Stanford Institute for Materials and Energy Sciences, SLAC National Accelerator Laboratory and Stanford University, Menlo Park, CA 94025, USA}
\affiliation{Geballe Laboratory for Advanced Materials, Stanford, CA 94305, USA}

\date{\today}

\begin{abstract}
Using determinant quantum Monte Carlo simulations, we compare charge transport in the two-dimensional attractive and repulsive Hubbard models at strong coupling, $|U|/t=6$. Although the interaction with opposite signs generates qualitatively different low-energy spin, charge, and pairing correlations, both exhibit approximately linear-in-temperature resistivity over a broad intermediate- and high-temperature regime. At asymptotically high temperature this common behavior follows from the moment expansion of the conductivity, whose leading contribution is even in $U$. More strikingly, the similarity persists to temperatures well below $|U|$, where strong interaction-dependent correlations have already developed, and also irrespective of whether resistivity crosses the MIR limit. Using the Nernst--Einstein relation, we find that the common linear-in-temperature resistivity is primarily associated with an approximately Curie-like charge compressibility and weakly temperature-dependent diffusivity. The two models separate only at lower temperatures, where the attractive model develops a pronounced feature in the charge diffusivity correlated with signatures of pair formation. These results show that linear-in-temperature incoherent transport in the Hubbard model can be remarkably insensitive to the microscopic nature of the low-energy correlations. Our results suggest that transport in the incoherent regime appears insensitive to what the system will ultimately become at low temperature; the distinction between competing low-energy states becomes visible only when their characteristic correlations acquire sufficiently long spatial or temporal coherence. 

\end{abstract}

\maketitle


\section{Introduction}
Understanding charge transport in strongly correlated materials remains a central challenge since strong interactions can invalidate the conventional quasiparticle description.\cite{gunnarsson_colloquium_2003,hussey_universality_2004,emery_superconductivity_1995,huang_strange_2019} In Fermi-liquid-like systems, as the temperature increases, the mean free path of charge-carrying quasiparticles approaches the scale of the Fermi wavelength $k_{F}^{-1}$, manifesting macroscopically as a gradual saturation of the resistivity, which is known as the Mott--Ioffe--Regel (MIR) condition, corresponding to a resistivity of $\rho = \hbar/e^{2}$ in a two-dimensional (2D) system.\cite{mott_conduction_1972,gunnarsson_colloquium_2003,hussey_universality_2004,werman_MIR_2016} 
Numerous studies have reported resistivities exceeding the MIR limit in a wide variety of metallic systems, including the normal states of cuprate high-$T_c$ superconductors, transition-metal dichalcogenides (TMDs), infinite-layer nickelates, heavy-fermion compounds, and magic-angle twisted bilayer graphene (MATBG).\cite{gurvitch_resistivity_1987,martin_normal-state_1990,takagi_systematic_1992,mackenzie_normal-state_1996,fournier_insulator-metal_1998,cooper_anomalous_2009,bruin_similarity_2013,legros_universal_2018,cao_strange_2020,jaoui_quantum_2022,lee_linear-in-temperature_2023,wei_linear_2024,xia_bandwidth-tuned_2026} These systems exhibit not only bad-metal behavior, with the resistivity increasing with temperature beyond the MIR limit\cite{emery_superconductivity_1995}, thus signaling the breakdown of the quasiparticle description, but also strange-metal behavior, characterized by a robust linear-in-temperature resistivity over a broad temperature range extending from below the characteristic phonon scale to temperatures at which the resistivity exceeds the MIR limit.\cite{phillips_stranger_2022,gunnarsson_colloquium_2003,hussey_universality_2004} While the linear-in-temperature resistivity appears universally in these materials with distinct ground states, its microscopic origins remain poorly understood, leading to a more fundamental question of how strange metallicity emerges initially: To what extent the phenomenon of linear-in-temperature resistivity is related to the low-energy correlations and the ground states, and to what extent it is instead a generic property of incoherent transport in a strongly correlated lattice system?

Linear-in-temperature resistivity can emerge from several different phenomenologies, including marginal-Fermi-liquid\cite{varma_phenomenology_1989,hlubina_resistivity_1995}, diffusion-based descriptions involving a universal bound on charge diffusivity,\cite{hartnoll_theory_2014} and an approximately Curie-like charge compressibility $\chi_c\propto 1/T$, accompanied by a weak temperature dependence of diffusivity.\cite{perepelitsky_transport_2016} Beyond the phenomenological explanations, the microscopic origin of strange metallicity in real materials remains unresolved, with no unified framework established across different material platforms. Numerical studies on microscopic models such as the repulsive Hubbard model, which is considered a promising pathway to study non-Fermi liquid behaviors and strange metals, have successfully demonstrated the strange-metal transport feature across a wide range of doping levels and temperatures.\cite{huang_strange_2019,brown_bad_2019} For systems with different low-energy correlations from the repulsive Hubbard model, the attractive Hubbard model is also reported recently to exhibit linear-in-temperature resistivity with perturbatively small $U/t$ in the weak-coupling limit.\cite{kiely_transport_2021}

Pinpointing the role of correlations in the incoherent transport is an important clue for understanding the overall underlying mechanisms. Building on the repulsive case studied before, the attractive Hubbard model (with the same interaction magnitude $\left| U\right|$) provides a clean comparison by changing the dominant low-energy correlations without altering the interaction-energy scale. Moreover, the attractive Hubbard model is free of the fermion sign problem of the determinantal Quantum Monte Carlo (DQMC) method\cite{blankenbecler_monte_1981,white_numerical_1989}, allowing access to substantially lower temperatures than are generally reachable in the doped repulsive model.

In this work, we use DQMC to study and compare the longitudinal charge transport of the strongly-correlated attractive and repulsive Hubbard models under a fixed magnitude $\left|U\right|$ of the on-site interaction. We find that the DC transport of both attractive and repulsive systems exhibits strange-metal behavior over a wide range of temperatures, despite the fundamentally different ground states of the two models. We demonstrate that the linear-in-temperature resistivity in the strongly coupled Hubbard model is remarkably insensitive to the sign of the interaction over an extended incoherent-temperature regime, despite the very different low-energy correlations generated by attractive and repulsive interactions. The commonality of the resistivity is irrespective of whether resistivity crosses the MIR limit. The common \(T\)-linear resistivity is primarily associated with an approximately Curie-like charge compressibility and weakly temperature-dependent diffusivity. The differences in the resistivities of the attractive and repulsive Hubbard models emerge at low temperature, which is attributed to the formation of pairs in the attractive case, and is illuminated by the charge diffusivity.

\section{Model and Methodology}
The Hubbard model is denoted by
\begin{equation}
    H=-t\sum_{ij,\sigma} c_{i\sigma}^{\dagger}c_{j\sigma}+U\sum_{i} \left( n_{i\uparrow}-\frac{1}{2} \right)\left( n_{i\downarrow}-\frac{1}{2} \right)-\mu\sum_{i,\sigma}n_{i\sigma},
\end{equation}
where $c_{i\sigma}$ and $c^{\dagger}_{i\sigma}$ are, respectively, the annihilation and creation operators of a single electron on the site $i$, with spin $\sigma$. For the scope of this paper, we consider only a uniform hopping $t$ between nearest-neighbor sites and a uniform on-site Coulomb interaction potential $U$. The magnitude and sign of the interaction term $U$ are key factors in regulating the intrinsic physical properties of the Hubbard model. Although the attractive and repulsive Hubbard models are related by a particle-hole transformation on one spin species, away from half filling the transformation interchanges chemical-potential and Zeeman-field sectors. Thus systems at the same density and $U$ of opposite signs are not generally symmetry-equivalent, making their transport comparison nontrivial. Unless specified otherwise, we use $U=\pm 6t$, on an $8\times 8$ square lattice with periodic boundaries throughout the paper.

For both attractive and repulsive Hubbard models, we investigate the DC charge transport properties by calculating the resistivity and conductivity, which are related to the unequal imaginary-time current-current correlator
\begin{equation}
    \Lambda_{xx} \left(\tau \right)=\frac{1}{N} \left\langle T_{\tau}J_{x}(\tau )J_{x}(0) \right\rangle
\end{equation}
via the Kubo formulas, where $\tau$ is the imaginary time. We employ the maximum entropy (MaxEnt) method to analytically continue $\Lambda_{xx}\left(\tau \right)$ to the real-frequency longitudinal conductivity $\sigma_{xx}\left(\omega\right)$, and the obtained conductivity $\sigma_{xx}\left(\omega\right)$ and DC resistivity $\rho_{xx}$ is validated with the low-temperature proxies.\cite{jaynes_information_1957,silver_maximum-entropy_1990,jarrell_bayesian_1996,huang_strange_2019} In the infinite-temperature limit, a moment expansion of $\sigma_{xx}$, derived from the short-time behavior of the real-time current-current correlator $\Lambda_{xx}(t)$, predicts $\sigma_{xx}\propto T^{-1}$ and hence a linear-in-temperature DC resistivity.\cite{huang_strange_2019} Detailed derivations and proofs about this high-temperature expansion and the low-temperature resistivity proxies are included in the supplementary materials.\cite{supplemental_material}

It should be noted that there is still a gap between the lowest temperature studied for the attractive case in this work and the expected superconducting order temperature, so our work is limited to the study of normal state transport properties, though the attractive Hubbard model is sign-free. Below the temperature of the onset of finite charge stiffness, the conductivity necessarily contains a nonzero zero-frequency $\delta$-function weight, which cannot be reliably distinguished from the regular component by the MaxEnt analytic continuation. Therefore, we conservatively restrict our DC resistivity analysis to temperatures above this scale. Discussion and derivations about the temperature cutoff for the attractive Hubbard model is included in the supplementary materials.\cite{supplemental_material}

\section{Results}

\begin{figure*}[!htbp]
    \centering
    \includegraphics[width=\linewidth]{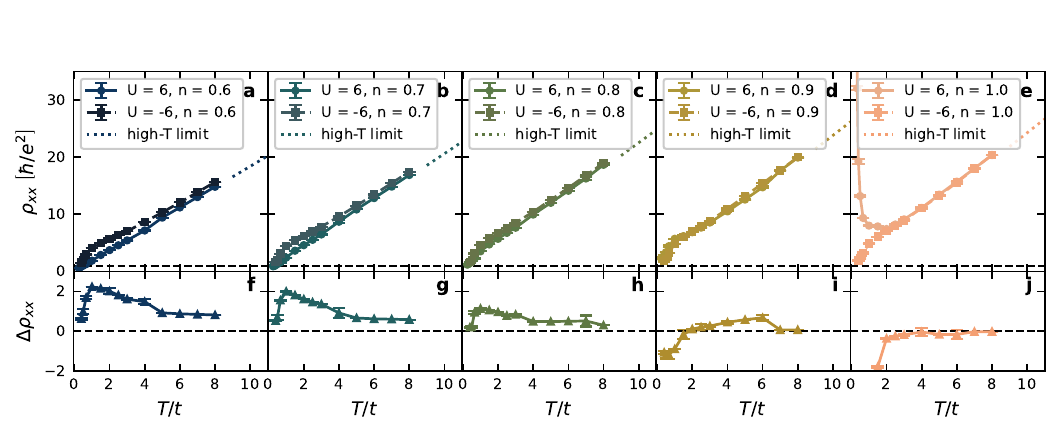}
    \caption{\textbf{a--e} Temperature dependence of the dc resistivity $\rho_{xx}$ in the attractive and repulsive Hubbard models, obtained from DQMC simulations combined with MaxEnt analytic continuation. The model parameters are $\lvert U\rvert/t=6$. The black dashed line marks the MIR limit, $\rho_{xx}=\hbar/e^2$, while the dotted lines show the high-temperature asymptotes obtained from the moments expansion. \textbf{f--j} Corresponding resistivity difference between the attractive and repulsive models, defined as $\Delta\rho_{xx}\left(\left|U\right|,n\right)\equiv\rho_{xx}(U<0,n)-\rho_{xx}(U>0,n)$.
}
    \label{fig:resistivity}
\end{figure*}

The difference in the ground states of attractive and repulsive Hubbard model raises the question of whether the strange-metal transport would be affected by the sign of the on-site interaction and the physical properties of the ground state. We therefore study the charge transport of the two systems. \cref{fig:resistivity} illustrates the DC longitudinal resistivity $\rho_{xx}\left(\omega=0\right)$ of both cases. As predicted by the moment expansion of the DC conductivity $\sigma_{xx}$ which is even in $U$, in the high-temperature limit where $T\gg \left|U\right|$, the temperature dependence of the resistivity in both cases converges to the same straight line passing through the origin.  Over a fairly wide temperature range from $T\gg \left|U\right|>t$ to $\left|U\right|>T\sim t$, both the attractive and repulsive Hubbard models exhibit the characteristic linear-in-temperature DC resistivity of strange metals. 

At half-filling, as shown in \cref{fig:resistivity}, the resistivities of the attractive and repulsive systems are almost identical within numerical precision when $T\gtrsim 2t$, and the difference between them does not become significant until the Mott gap opens and the resistivity experiences an upturn in the repulsive system. Away from half-filling, while the resistivities of both systems remain approximately linear in temperature at relatively high temperature $T\gtrsim 2t$, the difference in the slope of $\rho_{xx}\left(T\right)$ becomes significant as the hole doping increases. For all dopings, we observe a downturn in resistivity for the attractive Hubbard model, and we find a similar phenomenon in the resistivity of slightly hole-doped repulsive Hubbard systems. These features are robust to finite-size effects, as shown in \cref{SM-fig:resistivity_finite_size}.

\begin{figure}[!htbp]
    \centering
    \includegraphics[width=\linewidth]{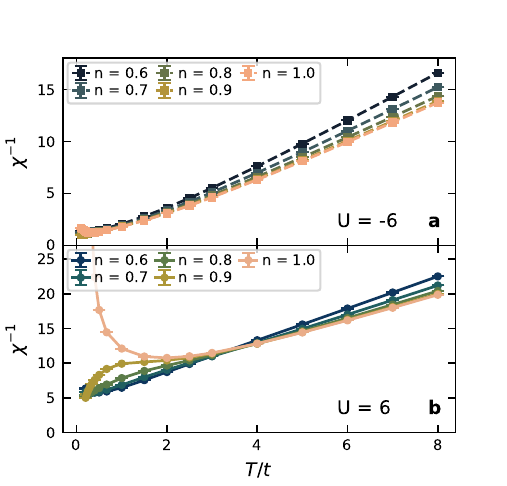}
    \caption{Inverse charge compressibility $\chi^{-1}$ of \textbf{a} the attractive Hubbard model, and \textbf{b} the repulsive Hubbard model, as functions of temperature $T=1/\beta$, obtained through DQMC simulations, with $\mid U\mid/t = 6$.}
    \label{fig:U6_U-6_inv_compressibility}
\end{figure}

\begin{figure}[!htbp]
    \centering
    \includegraphics[width=\linewidth]{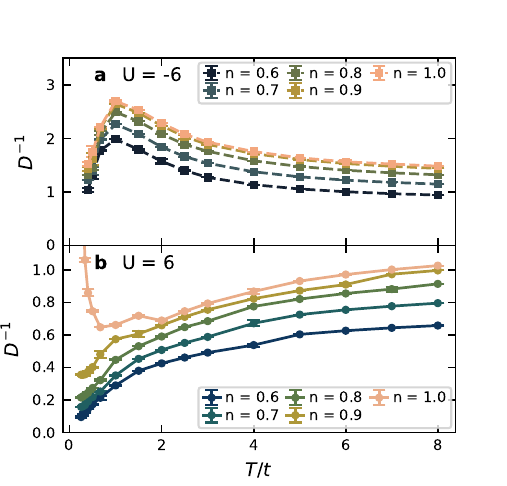}
    \caption{Inverse charge diffusivity $\mathcal{D}^{-1}$ of \textbf{a} the attractive Hubbard model, and \textbf{b} the repulsive Hubbard model, as functions of temperature $T=1/\beta$, obtained through DQMC simulations, with $\mid U\mid/t = 6$.}
    \label{fig:U6_U-6_inv_diffusivity}
\end{figure}

To illuminate the differences in the low temperature features of the repulsive and attractive Hubbard models, we examine the charge compressibility $\chi$ and the diffusivity $\mathcal{D}$, which are connected with the DC resistivity by the Nernst--Einstein relation,
\begin{equation}
    \rho_{xx}=\chi^{-1}\mathcal{D}^{-1}.
\end{equation}
As plotted in \cref{fig:U6_U-6_inv_compressibility}, the inverse charge compressibility $\chi^{-1}$ of both the attractive and repulsive Hubbard models away from half-filling exhibits an approximately linear temperature dependence at high temperatures and gradually saturates upon cooling. In the high-temperature regime where the resistivity is linear in temperature, \cref{fig:U6_U-6_inv_compressibility,fig:U6_U-6_inv_diffusivity} show that, in both the attractive and repulsive models, this linearity arises primarily from the approximate $1/T$ scaling of the charge compressibility $\chi$, while the charge diffusivity $\mathcal{D}$ remains nearly temperature independent. Upon cooling, the charge transport of both the attractive and repulsive Hubbard models experiences a crossover from compressibility-dominated to diffusivity-dominated, as supported by previous research.\cite{trivedi_deviations_1995,huang_strange_2019} The attractive Hubbard model exhibits a peak in the inverse diffusivity $\mathcal{D}^{-1}$ for all dopings at low temperature, leading to the low-temperature downturn seen in the DC resistivity. While this behavior is qualitatively distinct from the inverse diffusivity of the repulsive Hubbard model, a previous study demonstrated similar low-temperature features in the spin diffusivity $\mathcal{D}_{s}$ of the repulsive Hubbard model, which can be canonically mapped to the charge diffusivity of the attractive Hubbard model.\cite{ulaga_spin_2021} From \cref{fig:U6_U-6_inv_compressibility}, we note that similar low-temperature downturn in resistivity in slightly hole-doped repulsive Hubbard model with $\langle n\rangle = 0.9$ is attributed to the charge compressibility rather than the diffusivity as in the attractive case.

\begin{figure}[!htbp]
    \centering
    \includegraphics[width=\linewidth]{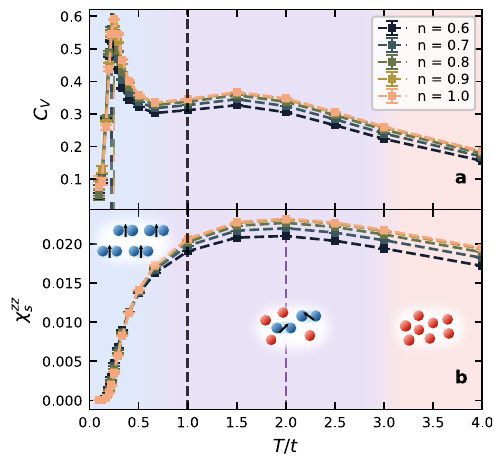}
    \caption{Temperature scales associated with pair formation in the attractive Hubbard model. \textbf{a} Specific heat $C_{V}$, and \textbf{b} uniform spin susceptibility $\chi_s^{zz}$ as functions of temperature $T=1/\beta$, calculated using DQMC for $U/t=-6$ at the indicated fillings. The black vertical dashed line marks the temperature at which the inverse diffusivity $\mathcal{D}^{-1}$ reaches its maximum. The purple vertical dashed line in \textbf{b} indicates the peak temperature of $\chi_s^{zz}$. In \textbf{a}, the colored vertical dashed lines mark the low-temperature peaks in $C_{V}$, with each color corresponding to the filling $\left\langle n\right\rangle$ of the associated data.}
    \label{fig:U-6_pairing_T}
\end{figure}

To investigate the origin of the low-temperature peak in the inverse diffusivity $\mathcal{D}^{-1}$ and the associated downturn in the resistivity of the attractive Hubbard model, we examine and plot the temperature dependence of the specific heat $C_{V}$ and uniform spin susceptibility 
\begin{equation}
    \chi^{zz}_{s}\left(\mathbf{Q}=0\right) = \frac{1}{N} \int_{0}^{\beta} d\tau \left< S^{z}_{\mathbf{Q}=0}\left( \tau \right) S^{z}_{\mathbf{Q}=0}\left( 0 \right) \right>
\end{equation}
in \cref{fig:U-6_pairing_T}, which can show the temperature scales related to pairing in the attractive Hubbard model. $C_{V}$ of the attractive Hubbard model shows a double-peak structure similar to the repulsive case reported in the literature.\cite{paiva_signatures_2001,duffy_specific_1997} The high-temperature peak in $C_V$ is associated with local-pair formation and quenching of spin excitations, whereas the low-temperature peak reflects the development of collective charge and pairing phase correlations.\cite{hurt_destruction_2005} The peak in the uniform spin susceptibility $\chi^{zz}_{s}\left(\mathbf{Q}=0\right)$, occurring near the temperature of the high-temperature peak in $C_{V}$, is consistent with the onset of pair formation, which is reported to occur at a higher temperature than the formation of quasi-long-range phase coherence under intermediate-to-strong coupling.\cite{fontenele_two-dimensional_2022} In the attractive Hubbard model, the concurrence of the thermodynamic and magnetic signatures suggests that $T\sim 2t$ represents a crossover scale related to preformed pairs. Decreasing from $T\sim 2t$, the suppression of $\chi^{zz}_{s}\left(\mathbf{Q}=0\right)$ agrees with the formation of spin-singlet pairs and the associated depletion of low-energy spin excitations. As shown in \cref{fig:U-6_pairing_T}, the temperature of the peak in $\mathcal{D}^{-1}$ which appears only in the attractive Hubbard model falls between the temperature scales of pair formation and phase correlation, raising the possibility that the peak in $\mathcal{D}^{-1}$ is associated with pair formation in the attractive Hubbard model.

\section{Discussion}
Through investigating the normal state transport properties of the attractive and repulsive Hubbard models, we find that the compressibility-controlled strange-metal transport, which manifests as a DC resistivity that remains linear in temperature throughout the incoherent regime over a broad temperature range extending down to $T\sim 2t$, is insensitive to what the system will ultimately become at low temperature; the distinction between competing low-energy states becomes visible only when their characteristic correlations acquire sufficiently long spatial or temporal coherence. The occurrence of similar transport phenomenology in models with different dominant low-energy correlations suggests that the observed linear resistivity may reflect transport constraints that are more general than any particular ordering tendency.

At low temperatures, the differences between attractive and repulsive Hubbard models appear in the diffusivity. In this paper, we propose a potential interpretation to the emergence of the low-temperature differences in the resistivities of the attractive and repulsive Hubbard models. In the attractive Hubbard model, the diffusivity-dominated transport behavior can be understood as a crossover between pair formation and many-body phase coherence as temperature is lowered. One possible interpretation is that pair formation initially suppresses charge diffusion, while the subsequent development of longer-lived collective pair correlations reverses this trend. While the initially incoherent Cooper pairs are forming, the thermodynamic charge response is preserved, but charge diffusion is simultaneously suppressed, resulting in a smaller diffusivity (and larger inverse diffusivity) when compared to the repulsive Hubbard model in the same temperature range. Establishing this mechanism will require a direct dynamical probe of pair correlations. Studies of a model system consisting of a 2D granular $\mathrm{Pb}$ film covered by an $\mathrm{Ag}$ overlayer have likewise revealed a clear separation between the development of a finite local superconducting order-parameter amplitude and the establishment of global superconducting phase coherence, with the former occurring before the latter.\cite{merchant_crossover_2001}

Since the resistivities of the attractive and repulsive cases converge at high temperatures to the same straight line passing through the origin, as predicted by the moments expansion, the downturn in resistivity of the attractive Hubbard model at low temperatures also leads to the difference in the slope of the resistivity versus temperature between the doped attractive and repulsive cases in the intermediate temperature region.

Further study of the pair density-density dynamical correlator $\left\langle D_{i}\left(\tau\right) D_{j}\left(0\right)\right\rangle$ would help clarify the influence of incoherent preformed pairs on charge diffusion. Furthermore, an important direction for future work is to determine whether systems with electron–phonon coupling exhibit charge transport behavior similar to that of the attractive Hubbard model. It would also be valuable to extend the calculations to different lattice geometries, which may stabilize distinct ground states and correlation patterns. Such studies would help distinguish universal transport features from effects specific to particular models and lattice geometries.

\section{Data Availability} Aggregated numerical data required to reproduce the figures in the main text can be found at \href{https://doi.org/10.5281/zenodo.22884589}{10.5281/zenodo.22884589}. Raw simulation data that support the findings of this study are stored on the Sherlock cluster at Stanford University and are available from the corresponding author upon reasonable request.

\section{Code Availability} The most up-to-date version of our DQMC simulation code, supporting simulations for both attractive and repulsive Hubbard model, can be accessed at \href{https://github.com/XineohpM/dqmc-dev}{https://github.com/XineohpM/dqmc-dev}.

\section{Acknowledgments} 
We are grateful for helpful discussions with Brian Moritz, Steven A. Kivelson, Wen O. Wang, and Evyatar Tulipman.

\bibliography{main}

\end{document}


\setcounter{secnumdepth}{2}
\setcounter{section}{0}


\begin{center}
    {\large\bfseries Supplemental Material for `` Robust Strange Metallicity across Attractive and Repulsive Hubbard Models''}
\end{center}
\section{Simulation Setup}
Determinant quantum Monte Carlo (DQMC) data reported in the main text of this paper are obtained from simulations performed using $4\times 10^{4}$ warm-up sweeps and $4\times 10^{5}$ measurement sweeps. $100$ to $400$ independently seeded Markov chains are run for each set of parameters. For all sets of parameters, the imaginary time step size satisfies $\Delta\tau/t\leq 0.05$ to reduce effects from Trotter error, where $t$ is the hopping term in the standard Hubbard Hamiltonian, and the number of imaginary time slices $L=\beta/\Delta \tau\geq 40$. All chemical potentials $\mu$ are fine-tuned so that the average particle density satisfies $\left|\left\langle n\right\rangle-n_{\text{target}}\right|<2\times 10^{-4}$. 

We perform a temperature-annealing procedure for the MaxEnt analytic continuation. The input model function is constructed from the infinite-temperature moment expansion. The resulting spectrum is then used as the default model for the next lower temperature, and this procedure is repeated sequentially as the temperature decreases.\cite{aryanpour_dynamical_2006,huang_strange_2019,wang_the_2023} The ``BT" method is used to select the hyper-parameter $\alpha$ for the maximum entropy (MaxEnt) analytical continuation process.\cite{bergeron_algorithms_2016}

\section{Detailed Method}
The formalism of the derivations related to charge transport and the conventions of sign, normalization and notations used in this work are listed below for consistence and completeness.

Consider the 2D square-lattice Hubbard model with nearest-neighbor hopping $t$,
\begin{equation}
    H=Ud+K_x+K_y-\mu N,
    \label{hubbard_hamiltonian}
\end{equation}
where
\begin{equation}
    K_{x}=-t\sum_{i\sigma}\left(c^\dagger_{i+\hat x,\sigma}c_{i,\sigma}+c^\dagger_{i,\sigma}c_{i+\hat x,\sigma}\right)
\end{equation}
and
\begin{equation}
    K_{y}=-t\sum_{i\sigma}\left(c^\dagger_{i+\hat y,\sigma}c_{i,\sigma}+c^\dagger_{i,\sigma}c_{i+\hat y,\sigma}\right).
\end{equation}
For a spatially uniform vector potential $\mathbf{A}=\left(A_{x},0\right)$ along the $x$ direction, the kinetic term
\begin{equation}
    K_{x}\left( A_{x} \right) =-t\sum_{i\sigma} \left( e^{-iA_{x}}c_{i+\hat{x} ,\sigma}^{\dagger}c_{i,\sigma}+e^{iA_{x}}c_{i,\sigma}^{\dagger}c_{i+\hat{x} ,\sigma} \right),
\end{equation}
which can be expanded in terms of $A_{x}$ as
\begin{equation}
    K_{x}\left( A_{x} \right) =K_{x}\left( A_{x}=0 \right) -A_{x}J_{x}+\frac{1}{2} A_{x}^{2}\tau_{xx} +O\left( A_{x}^{3} \right),
\end{equation}
where the current operator
\begin{align}
    J_{x}&=-\left.\frac{\partial H}{\partial A_{x}} \right|_{A_{x} =0} \\
    &= -it\sum_{i\sigma} \left( c_{i+\hat{x} ,\sigma}^{\dagger}c_{i,\sigma}-c_{i,\sigma}^{\dagger}c_{i+\hat{x} ,\sigma} \right),
\end{align}
and the diamagnetic operator
\begin{align}
    \tau_{xx} &=\left. \frac{\partial^{2} H}{\partial A_{x}^{2}} \right|_{A_{x} =0} \\
    &=t\sum_{i\sigma}\left(c^\dagger_{i+\hat x,\sigma}c_{i,\sigma}+c^\dagger_{i,\sigma}c_{i+\hat x,\sigma}\right) \\
    &=-K_{x}\left(A_{x}=0\right),
\end{align}
following a unit convention of the reduced Planck constant, the magnitude of the charge of electron and the lattice constant satisfying $\hbar=e=a=1$. Therefore, the Kubo formula of the longitudinal optical conductivity is written as\cite{ding_intrinsic_2025}
\begin{equation}
    \sigma_{xx}\left(\omega\right)=\frac{i}{\omega+i0^{+}}\left[\chi^{R}_{xx}\left(\omega\right)+\frac{\left\langle\tau_{xx}\right\rangle}{V}\right]=\frac{i}{\omega+i0^{+}}\left[\chi^{R}_{xx}\left(\omega\right)-\frac{\left\langle K_{x}\right\rangle}{V}\right].
    \label{conductivity_Kubo}
\end{equation}

The Fourier transform of $\mathbf A$ consists only of the zero-momentum $\mathbf q = 0$ component due to its uniformity in space. The vector potential $\mathbf A$ here is equivalent to an overall boundary twist $\Phi_{x}$ by
\begin{equation}
    A_{x}=\frac{\Phi_{x}}{L_{x}}.
\end{equation}
The isothermal charge stiffness $\bar{D}$ is defined as 
\begin{align}
    \bar{D}&=\frac{1}{V}\left.\frac{\partial^{2}F}{\partial A_{x}^{2}}\right|_{A_{x}=0} \\
    &=\frac{1}{V}\left\langle \left.\frac{\partial^{2}H}{\partial A_{x}^{2}}\right|_{A_{x}=0}\right\rangle-\frac{1}{V}\int^{\beta}_{0}d\tau\left[\left\langle \left.\frac{\partial H}{\partial A_{x}}\right|_{A_{x}=0}\left(\tau\right)\left.\frac{\partial H}{\partial A_{x}}\left(0\right)\right|_{A_{x}=0} \right\rangle - \left\langle \left.\frac{\partial H}{\partial A_{x}} \right|_{A_{x}=0} \right\rangle^{2}\right] \\
    &=-\frac{\left\langle K_{x}\right\rangle}{V}-\frac{1}{V}\int^{\beta}_{0}d\tau\left[\left\langle J_{x}\left(\tau\right) J_{x}\left(0\right)\right\rangle - \left\langle J_{x}\right\rangle^2\right],
    \label{isothermal_charge_stiffness}
\end{align}
where $F$ is the free energy.

\subsection{The isothermal and adiabatic charge stiffness}
In this section we provide a derivation of the isothermal and adiabatic charge stiffness, which limits the lowest temperature at which we can reliably measure the charge transport properties. The following derivation applies to systems of finite size at finite temperatures.

It is given that under finite temperature
\begin{align}
    \left\langle\left[J_{x}\left(t\right),J_{x}\left(0\right)\right]\right\rangle_{\beta}&=\left\langle J_{x}\left(t\right)J_{x}\left(0\right)\right\rangle_{\beta} - \left\langle J_{x}\left(0\right)J_{x}\left(t\right)\right\rangle_{\beta} \\
    &= \frac{1}{Z}\sum_{mn}\left(e^{-\beta E_{n}}-e^{-\beta E_{m}}\right) e^{i\left(E_{n}-E_{m}\right)t}\left|\left\langle m\left|J_{x}\right|n\right\rangle\right|^{2},
\end{align}
the retarded current-current correlator
\begin{align}
    \chi^{R}_{xx}\left(\omega\right)&=-\frac{i}{V}\int^{\infty}_{0}dte^{i\left(\omega+i0^{+}\right)t}\left\langle\left[J_{x}\left(t\right),J_{x}\left(0\right)\right]\right\rangle_{\beta} \\
    &=\frac{1}{ZV}\sum_{mn}\left(e^{-\beta E_{n}}-e^{-\beta E_{m}}\right)\frac{\left|\left\langle m\left|J_{x}\right|n\right\rangle\right|^{2}}{\omega+i0^{+}+E_{n}-E_{m}} \\
    &=\frac{1}{ZV}\sum_{mn}\left(e^{-\beta E_{n}}-e^{-\beta E_{m}}\right)\left|\left\langle m\left|J_{x}\right|n\right\rangle\right|^{2}\left[\mathcal{P}\left(\frac{1}{\omega+E_{n}-E_{m}}\right)-i\pi\delta\left(\omega+E_{n}-E_{m}\right)\right].
\end{align}
The imaginary part of the retarded current–current correlator has the Lehmann representation
\begin{equation}
    \mathrm{Im}\ \chi^{R}_{xx}\left(\omega\right)=-\frac{\pi}{ZV}\sum_{mn}\left(e^{-\beta E_{n}}-e^{-\beta E_{m}}\right)\left|\left\langle m\left|J_{x}\right|n\right\rangle\right|^{2}\delta\left(\omega+E_{n}-E_{m}\right),
    \label{im_retarded_corr_lehmann}
\end{equation}
which is odd in frequency and therefore vanishes at $\omega=0$. According to \cref{conductivity_Kubo}, the real part of the conductivity
\begin{align}
    \mathrm{Re}\ \sigma_{xx}\left(\omega\right)&=\mathrm{Re}\ \frac{i}{\omega+i0^{+}}\left[\chi^{R}_{xx}\left(\omega\right)-\frac{\left\langle K_{x}\right\rangle}{V}\right] \\
    &=\mathrm{Re}\ \left[i\mathcal{P}\left(\frac{1}{\omega}\right)+\pi\delta\left(\omega\right) \right]\left[\chi^{R}_{xx}\left(\omega\right)-\frac{\left\langle K_{x}\right\rangle}{V} \right] \\
    &=-\mathcal{P}\left(\frac{1}{\omega}\right)\mathrm{Im}\ \chi^{R}_{xx}\left(\omega\right)+\pi\delta\left(\omega\right)\left[\mathrm{Re}\ \chi^{R}_{xx}\left(\omega\right)-\frac{\left\langle K_{x}\right\rangle}{V} \right] \\
    &=-\frac{1}{\omega}\mathrm{Im}\ \chi^{R}_{xx}\left(\omega\neq 0\right)+\pi D\delta\left(\omega\right),
\end{align}
where $D$ is the finite-size Drude weight, also called the adiabatic charge stiffness, and $\pi D$ is the coefficient of the $0$-frequency delta function.\cite{shastry_sum_2006,mukerjee_signature_2008} At non-zero frequencies, according to \cref{im_retarded_corr_lehmann}, the real part of the regular conductivity has the Lehmann representation
\begin{align}
    \mathrm{Re}\ \sigma^{\mathrm{reg}}_{xx}\left(\omega\neq 0\right)&=-\frac{1}{\omega}\mathrm{Im}\ \chi^{R}_{xx}\left(\omega \right) \\
    &=\frac{\pi}{\omega ZV}\sum_{mn}\left(e^{-\beta E_{n}}-e^{-\beta E_{m}}\right)\left|\left\langle m\left|J_{x}\right|n\right\rangle\right|^{2}\delta\left(\omega+E_{n}-E_{m}\right).
    \label{re_reg_cond}
\end{align}

In the finite-temperature DQMC, we evaluate the imaginary-time current-current correlation function
\begin{equation}
    \Lambda_{xx}\left(\tau\right)=\frac{1}{V}\left\langle T_{\tau}J_{x}\left(\tau\right)J_{x}\left(0\right)\right\rangle=\frac{1}{VZ}\sum_{mn}e^{-\beta E_{n}}e^{-\tau\left(E_{m}-E_{n}\right)}\left|\left\langle m\left|J_{x}\right|n\right\rangle\right|^{2},
\end{equation}
which can be decomposed into the sum of
\begin{equation}
    \Lambda_{xx}^{E_{m}\neq E_{n}}=\frac{1}{VZ}\sum_{E_{m}\neq E_{n}}e^{-\beta E_{n}}e^{-\tau\left(E_{m}-E_{n}\right)}\left|\left\langle m\left|J_{x}\right|n\right\rangle\right|^{2}
\end{equation}
and a $\tau$-independent zero-frequency ballistic term
\begin{equation}
    C_{0}=\frac{1}{VZ}\sum_{E_{m}=E_{n}}e^{-\beta E_{n}}\left|\left\langle m\left|J_{x}\right|n\right\rangle\right|^{2}.
\end{equation}
According to \cref{im_retarded_corr_lehmann} and \cref{re_reg_cond}, $\Lambda_{xx}^{E_{m}\neq E_{n}}$ is related to the regular conductivity by
\begin{equation}
    \Lambda_{xx}^{E_{m}\neq E_{n}}=\int_{0^{+}}^{\infty} \frac{d\omega}{\pi} K_{B}\left( \tau ,\omega \right) \mathrm{Re} \  \sigma_{xx}^{\mathrm{reg}} \left( \omega \right),
\end{equation}
where the symmetric bosonic kernel
\begin{equation}
    K_{B}\left(\tau,\omega\right)=\frac{\omega\left[e^{\left(\beta-\tau\right)\omega}+e^{\tau\omega}\right]}{e^{\beta\omega}-1}.
\end{equation}
Since $\int^{\frac{\beta}{2}}_{0}d\tau K_{B}\left(\tau,\omega\right)= 1 $, the integral on the half interval from $0$ to $\beta/2$ of the imaginary-time current-current correlation function is
\begin{equation}
    \int^{\frac{\beta}{2}}_{0}d\tau\Lambda_{xx}\left(\tau\right)=\int^{\frac{\beta}{2}}_{0}d\tau C_{0}+\int^{\frac{\beta}{2}}_{0}d\tau\Lambda_{xx}^{E_{m}\neq E_{n}}=\frac{\beta C_{0}}{2}+\int^{\infty}_{0^{+}}\frac{d\omega}{\pi}\mathrm{Re}\ \sigma^{\mathrm{reg}}_{xx}\left(\omega\right).
\end{equation}
Since $\left\langle J_{x}\right\rangle=0$ due to the spatial inversion symmetry in our case, the isothermal charge stiffness $\bar{D}$ can be calculated by
\begin{align}
    \bar{D}&=-\frac{\left\langle K_{x}\right\rangle}{V}-\int^{\beta}_{0}d\tau \Lambda_{xx}\left(\tau\right)\\
    &=-\frac{\left\langle K_{x}\right\rangle}{V}-\frac{1}{VZ}\int^{\beta}_{0}d\tau \sum_{mn}e^{-\beta E_{n}}e^{-\tau\left(E_{m}-E_{n}\right)}\left|\left\langle m\left|J_{x}\right|n\right\rangle\right|^{2}.
\end{align}
The finite-size Drude weight $D$ can be calculated by
\begin{align}
    D&=-\frac{\left\langle K_{x}\right\rangle}{V}+\lim_{\omega\to 0^{+}}\mathrm{Re}\ \chi^{R}_{xx}\left(\omega\right) \\
    &=-\frac{\left\langle K_{x}\right\rangle}{V}-\frac{2}{\pi}\int^{\infty}_{0^{+}}d\omega \mathrm{Re}\ \sigma^{\mathrm{reg}}_{xx}\left(\omega\right) \\
    &=-\frac{\left\langle K_{x}\right\rangle}{V}-\int^{\beta}_{0}d\tau \Lambda^{E_{m}\neq E_{n}}_{xx}\left(\tau\right) \\
    &=-\frac{\left\langle K_{x}\right\rangle}{V}-\frac{1}{VZ}\int^{\beta}_{0}d\tau \sum_{E_{m}\neq E_{n}}e^{-\beta E_{n}}e^{-\tau\left(E_{m}-E_{n}\right)}\left|\left\langle m\left|J_{x}\right|n\right\rangle\right|^{2} \\
    &=-\frac{\left\langle K_{x}\right\rangle}{V}-\frac{1}{VZ}\sum_{E_{m}\neq E_{n}}\frac{e^{-\beta E_{m}}-e^{-\beta E_{n}}}{E_{n}-E_{m}}\left|\left\langle m\left|J_{x}\right|n\right\rangle\right|^{2},
\end{align}
and therefore\cite{shastry_sum_2006,mukerjee_signature_2008}
\begin{equation}
    \bar{D}=D-\beta C_{0}.
\end{equation}

In the present implementation, MaxEnt is performed on a strictly positive-frequency grid and contains no explicit $\delta\left(\omega\right)$ basis. It therefore cannot reconstruct an exact zero-frequency delta function, and the $\tau$-independent $C_{0}$ contribution contained in the imaginary-time correlator may be represented as unresolved low-frequency spectral weight, whereas the contribution associated with the finite-size isothermal stiffness $\bar{D}$ is absent from $\Lambda_{xx}(\tau)$ altogether. Since the MaxEnt spectrum is normalized to the imaginary-time current–current correlator integral, the temperature at which the integrated optical conductivity $\left(2/\pi\right)\sum_{i}\sigma_{\mathrm{ME}}\left(\omega_{i}\right)\Delta\omega_{i}$ falls short of the kinetic energy sum should coincide with the temperature at which the imaginary-time correlator integral $\int^{\beta}_{0}d\tau \Lambda_{xx}$ shows the same shortfall within the margin of error. A positive stiffness $\bar{D}>0$, rigorously implies a finite-size zero-frequency delta weight with $D=\bar{D}+\beta C_{0}>0$. Correspondingly, the smooth MaxEnt spectrum misses a one-sided spectral weight relative to the complete optical sum rule. In this regime, a finite DC resistivity inferred from the lowest-frequency MaxEnt value should not be interpreted as the total physical DC resistivity of the finite-size system. Thus, a significantly shortfall is a sufficient warning that the MaxEnt DC resistivity is unreliable, but it is not a necessary condition: even when the imaginary-time correlator integral agrees with the kinetic-energy sum, a nonzero $D=\beta C_{0}$ or an unresolved narrow regular peak may still make the DC extrapolation unreliable. It should be noted that the isothermal and adiabatic charge stiffness $\bar{D}$ and $D$ here are both results obtained for finite-size system, therefore the fact that $\bar{D} > 0$ does not imply that the system has entered a superconducting state.

For the attractive Hubbard model, we calculate the difference between the imaginary-time correlator integral and the kinetic energy sum, that is, the isothermal charge stiffness $\bar{D}$, under different temperatures, as listed in \cref{tab:charge_stiffness}. We use the lowest sampled temperature at which the relative error of the kinetic energy sum and the imaginary-time correlator integral is less than $2\%$ and $\left|\bar{D}\right|/t < 0.01$ as the lowest cutoff temperature for our MaxEnt analytical continuation. For the attractive Hubbard model with $U/t = -6$, the empirical cutoff temperature is $T/t = 0.4$ for all tested dopings.

\begin{table}[!htbp]
    \centering
    \resizebox{\linewidth}{!}{%
    \begin{tabular}{c c c c c c}
        \toprule
        & \multicolumn{5}{c}{Average filling $\left\langle n\right\rangle$} \\
        \cmidrule(lr){2-6}
        $\beta t$ & $0.6$ & $0.7$ & $0.8$ & $0.9$ & $1.0$ \\
        \midrule
        10 & $0.4767\pm5.519\times10^{-4}$ & $0.4979\pm4.818\times10^{-4}$
             & $0.5009\pm5.706\times10^{-4}$ & $0.4655\pm5.680\times10^{-4}$ & $0.3330\pm5.097\times10^{-4}$ \\
        8 & $0.4646\pm6.140\times10^{-4}$ & $0.4840\pm6.086\times10^{-4}$
             & $0.4853\pm7.824\times10^{-4}$ & $0.4379\pm6.349\times10^{-4}$ & $0.3216\pm1.009\times10^{-3}$ \\
        6 & $0.3906\pm5.977\times10^{-4}$ & $0.4135\pm6.283\times10^{-4}$
             & $0.4071\pm4.532\times10^{-4}$ & $0.3549\pm4.600\times10^{-4}$ & $0.2778\pm4.896\times10^{-4}$ \\
        5 & $0.2693\pm3.601\times10^{-4}$ & $0.2925\pm5.616\times10^{-4}$
             & $0.2884\pm4.797\times10^{-4}$ & $0.2515\pm6.727\times10^{-4}$ & $0.2122\pm4.439\times10^{-4}$ \\
        4.5 & $0.1798\pm5.521\times10^{-4}$ & $0.1997\pm3.924\times10^{-4}$
             & $0.1992\pm3.843\times10^{-4}$ & $0.1756\pm6.001\times10^{-4}$ & $0.1569\pm3.429\times10^{-4}$ \\
        4 & $0.09387\pm2.437\times10^{-4}$ & $0.1094\pm2.609\times10^{-4}$
             & $0.1111\pm3.237\times10^{-4}$ & $0.1022\pm3.779\times10^{-4}$ & $0.09455\pm3.406\times10^{-4}$ \\
        3.5 & $0.03711\pm1.959\times10^{-4}$ & $0.04597\pm2.156\times10^{-4}$
             & $0.04804\pm1.898\times10^{-4}$ & $0.04633\pm4.094\times10^{-4}$ & $0.04366\pm2.529\times10^{-4}$ \\
        3 & $0.01053\pm1.967\times10^{-4}$ & $0.01475\pm1.525\times10^{-4}$
             & $0.01673\pm1.423\times10^{-4}$ & $0.01585\pm1.467\times10^{-4}$ & $0.01483\pm1.774\times10^{-4}$ \\
        2.5 & $1.621\times10^{-3}\pm9.788\times10^{-5}$ & $4.056\times10^{-3}\pm1.201\times10^{-4}$
             & $5.484\times10^{-3}\pm1.957\times10^{-4}$ & $4.696\times10^{-3}\pm1.106\times10^{-4}$ & $4.958\times10^{-3}\pm3.120\times10^{-4}$ \\
        2 & $5.560\times10^{-4}\pm6.820\times10^{-5}$ & $1.309\times10^{-3}\pm7.025\times10^{-5}$
             & $1.942\times10^{-3}\pm6.916\times10^{-5}$ & $2.103\times10^{-3}\pm7.406\times10^{-5}$ & $2.067\times10^{-3}\pm6.369\times10^{-5}$ \\
        1.5 & $3.319\times10^{-4}\pm4.167\times10^{-5}$ & $5.375\times10^{-4}\pm3.778\times10^{-5}$
             & $8.020\times10^{-4}\pm4.066\times10^{-5}$ & $9.615\times10^{-4}\pm4.488\times10^{-5}$ & $1.078\times10^{-3}\pm3.853\times10^{-5}$ \\
        1 & $2.372\times10^{-4}\pm1.946\times10^{-5}$ & $2.813\times10^{-4}\pm2.236\times10^{-5}$
             & $3.342\times10^{-4}\pm2.494\times10^{-5}$ & $3.678\times10^{-4}\pm1.951\times10^{-5}$ & $4.105\times10^{-4}\pm1.984\times10^{-5}$ \\
        0.666667 & $1.162\times10^{-4}\pm1.009\times10^{-5}$ & $9.512\times10^{-5}\pm1.305\times10^{-5}$
             & $1.521\times10^{-4}\pm1.213\times10^{-5}$ & $1.558\times10^{-4}\pm1.353\times10^{-5}$ & $1.567\times10^{-4}\pm1.293\times10^{-5}$ \\
        0.5 & $5.315\times10^{-5}\pm7.705\times10^{-6}$ & $7.752\times10^{-5}\pm7.523\times10^{-6}$
             & $7.276\times10^{-5}\pm8.012\times10^{-6}$ & $7.216\times10^{-5}\pm7.985\times10^{-6}$ & $8.757\times10^{-5}\pm8.473\times10^{-6}$ \\
        0.4 & $4.511\times10^{-5}\pm6.008\times10^{-6}$ & $3.400\times10^{-5}\pm5.728\times10^{-6}$
             & $3.303\times10^{-5}\pm6.176\times10^{-6}$ & $4.151\times10^{-5}\pm6.360\times10^{-6}$ & $3.664\times10^{-5}\pm5.748\times10^{-6}$ \\
        0.333333 & $2.296\times10^{-5}\pm4.633\times10^{-6}$ & $1.946\times10^{-5}\pm4.258\times10^{-6}$
             & $2.384\times10^{-5}\pm3.853\times10^{-6}$ & $3.123\times10^{-5}\pm4.134\times10^{-6}$ & $2.507\times10^{-5}\pm4.223\times10^{-6}$ \\
        0.25 & $1.535\times10^{-5}\pm2.861\times10^{-6}$ & $1.082\times10^{-5}\pm2.913\times10^{-6}$
             & $1.428\times10^{-5}\pm2.707\times10^{-6}$ & $8.974\times10^{-6}\pm2.601\times10^{-6}$ & $1.131\times10^{-5}\pm2.643\times10^{-6}$ \\
        0.2 & $3.837\times10^{-6}\pm2.173\times10^{-6}$ & $5.474\times10^{-6}\pm2.286\times10^{-6}$
             & $3.468\times10^{-6}\pm1.877\times10^{-6}$ & $7.908\times10^{-6}\pm1.645\times10^{-6}$ & $4.993\times10^{-6}\pm1.846\times10^{-6}$ \\
        0.166667 & $4.877\times10^{-7}\pm1.662\times10^{-6}$ & $5.145\times10^{-6}\pm1.460\times10^{-6}$
             & $5.169\times10^{-6}\pm1.622\times10^{-6}$ & $1.522\times10^{-6}\pm1.285\times10^{-6}$ & $7.683\times10^{-6}\pm1.286\times10^{-6}$ \\
        0.142857 & $1.667\times10^{-6}\pm1.535\times10^{-6}$ & $3.228\times10^{-6}\pm1.433\times10^{-6}$
             & $2.441\times10^{-6}\pm1.205\times10^{-6}$ & $4.403\times10^{-6}\pm1.216\times10^{-6}$ & $1.173\times10^{-6}\pm9.652\times10^{-7}$ \\
        0.125 & $-5.939\times10^{-8}\pm1.370\times10^{-6}$ & $3.669\times10^{-6}\pm1.082\times10^{-6}$
             & $5.701\times10^{-7}\pm9.504\times10^{-7}$ & $1.119\times10^{-6}\pm8.495\times10^{-7}$ & $2.220\times10^{-6}\pm9.173\times10^{-7}$ \\
        \bottomrule
    \end{tabular}
    }
    \caption{Isothermal charge stiffness $\bar{D}$ of the attractive Hubbard model under different temperatures, with $U/t = -6$, $\left\langle n\right\rangle = 0.6, 0.7, 0.8, 0.9$ and $1.0$.}
    \label{tab:charge_stiffness}
\end{table}

\subsection{Infinite-temperature series expansion}
The infinite-temperature series expansion used in this work follows that described in the Supplemental Material accompanying Ref.~\cite{huang_strange_2019}. In this section, we prove that the infinite-temperature series expansion is even in the on-site interaction term $U$ with fixed filling $n$.

Now consider the real-time current-current correlation function
\begin{equation}
    \Lambda_{xx}\left(t\right)=\frac{1}{V}\left\langle J_{x}\left(t\right)J_{x}\left(0\right)\right\rangle=\frac{1}{VZ}\sum_{mn}e^{-\beta E_{m}}e^{i\left(E_{m}-E_{n}\right)t}\left|\left\langle m\left|J_{x}\right| n\right\rangle\right|^{2}.
\end{equation}
Under the limit of infinite temperature, we have
\begin{equation}
    \lim_{T\rightarrow \infty} \Lambda_{xx} \left( t \right) =\lim_{T\rightarrow \infty} 2T\int_{-\infty}^{\infty} \frac{d\omega}{2\pi} e^{-i\omega t}\mathrm{Re} \  \sigma_{xx} \left( \omega \right),
    \label{inf_T_limit}
\end{equation}
which can be expanded near $t=0$ as
\begin{equation}
    \lim_{T\rightarrow \infty} \Lambda_{xx} \left( t \right) =\lim_{T\rightarrow \infty} 2T\sum_{k=0}^{\infty} \frac{\left( -it \right)^{k}}{k!} \int_{-\infty}^{\infty} \frac{d\omega}{2\pi} \omega^{k} \mathrm{Re} \  \sigma_{xx} \left( \omega \right)=\lim_{T\rightarrow \infty} 2T\sum_{k=0}^{\infty} \frac{\left( -it \right)^{k} \mu_{k}}{k!},
\end{equation}
where the $\text{k}^{\text{th}}$-moment of the optical conductivity 
\begin{equation}
    \mu_{k}=\int_{-\infty}^{\infty} \frac{d\omega}{2\pi} \omega^{k} \mathrm{Re} \ \sigma_{xx} \left( \omega \right).
\end{equation}
Since $\mathrm{Re}\ \sigma_{xx}\left(\omega\right)$ is even in $\omega$ according to \cref{re_reg_cond}, all odd-order moments vanish. The even-order moments
\begin{align}
    \mu_{2k} &=\int_{-\infty}^{\infty} \frac{d\omega}{2\pi} \omega^{2k} \mathrm{Re} \ \sigma_{xx} \left( \omega \right) \\ &=\frac{1}{2T} \left. \left( i\frac{d}{dt} \right)^{2k} \Lambda_{xx} \left( t \right) \right\vert_{t=0} \\
    &=\frac{1}{2TV} \left< \left( \mathcal{L}^{2k} J_{x} \right) J_{x} \right>,
\end{align}
where $\mathcal{L}$ is the Liouvillian, such that
\begin{equation}
    \mathcal{L}J_{x}=\left[H,J_{x}\right]=\left[Ud+K,J_{x}\right]
\end{equation}
since $\left[N,J_{x}\right]=0$. Therefore, it is sufficient to show that $\mu_{2k}$ is even in $U$ for all integer $k$.

Define a unitary transformation $\mathcal{G}$, such that
\begin{equation}
    \mathcal{G}c_{i\sigma}\mathcal{G}^{\dagger}=\eta_{i}c_{i\sigma},\ \mathcal{G}c_{i\sigma}^{\dagger}\mathcal{G}^{\dagger}=\eta_{i}c_{i\sigma}^{\dagger},
\end{equation}
where the staggered phase factor on the bipartite square lattice
\begin{equation}
    \eta_{i} =\begin{cases}+1&\text{, if } i\in A,\\ -1&\text{, if } i\in B.\end{cases}
\end{equation}
Then we have
\begin{equation}
    \mathcal{G}J_{x}\mathcal{G}^{\dagger}=-it\mathcal{G}\sum_{i\sigma} \left( c_{i+\hat{x} ,\sigma}^{\dagger}c_{i,\sigma}-c_{i,\sigma}^{\dagger}c_{i+\hat{x} ,\sigma} \right)\mathcal{G}^{\dagger}=it\sum_{i\sigma}\left( c_{i+\hat{x} ,\sigma}^{\dagger}c_{i,\sigma}-c_{i,\sigma}^{\dagger}c_{i+\hat{x} ,\sigma} \right)=-J_{x},
\end{equation}
\begin{equation}
    \mathcal{G}K\mathcal{G}^{\dagger}=-t\mathcal{G}\sum_{\left\langle i,j\right\rangle\sigma}\left(c_{i\sigma}^{\dagger}c_{j\sigma}+c_{j\sigma}^{\dagger}c_{i\sigma}\right)\mathcal{G}^{\dagger}=t\sum_{\left\langle i,j\right\rangle\sigma}\left(c_{i\sigma}^{\dagger}c_{j\sigma}+c_{j\sigma}^{\dagger}c_{i\sigma}\right)=-K,
\end{equation}
and 
\begin{equation}
    \mathcal{G}Ud\mathcal{G}^{\dagger}=U\mathcal{G}\sum_{i}n_{i\uparrow}n_{i\downarrow}\mathcal{G}^{\dagger}=Ud.
\end{equation}
Therefore,
\begin{equation}
    \mathcal{G}\left(Ud+K\right)\mathcal{G}^{\dagger}=Ud-K=-\left[\left(-U\right)d+K\right].
\end{equation}
Since $\left< O \right>_{\lambda =\mu \beta} =\frac{\mathrm{Tr} \left( e^{\lambda N}O \right)}{\mathrm{Tr} \left( e^{\lambda N} \right)} =\left< \mathcal{G} O\mathcal{G}^{\dagger} \right>_{\lambda}$ for any operator $O$ under the infinite-temperature limit in the grand canonical ensemble, we have
\begin{align}
    \mu_{2k}\left(U\right)&=\frac{1}{2TV}\left\langle\left[\mathcal{L}^{2k}\left(U\right)J_{x}\right]J_{x}\right\rangle \\
    &=\frac{1}{2TV}\left\langle\mathcal{G}\left[\mathcal{L}^{2k}\left(U\right)J_{x}\right]J_{x}\mathcal{G}^{\dagger}\right\rangle \\
    &=\frac{1}{2TV}\left\langle\left( -1 \right)^{2k+1} \mathcal{L}^{2k} \left( -U \right) J_{x}\left( -J_{x} \right)\right\rangle \\
    &=\mu_{2k}\left(-U\right),
\end{align}
which leads to the conclusion that $\Lambda_{xx}\left(t\right)$ and $T\mathrm{Re}\ \sigma_{xx}$ are both even in $U$ with fixed filling $n$ under the infinite-temperature limit. The real-frequency dependence of $T\mathrm{Re}\ \sigma_{xx}$ is plotted in \cref{fig:high_T_Tsigma}.

\section{Supplemental Figures}

\subsection{Average sign}
\begin{figure}[!htbp]
    \centering
    \includegraphics[width=0.5\linewidth]{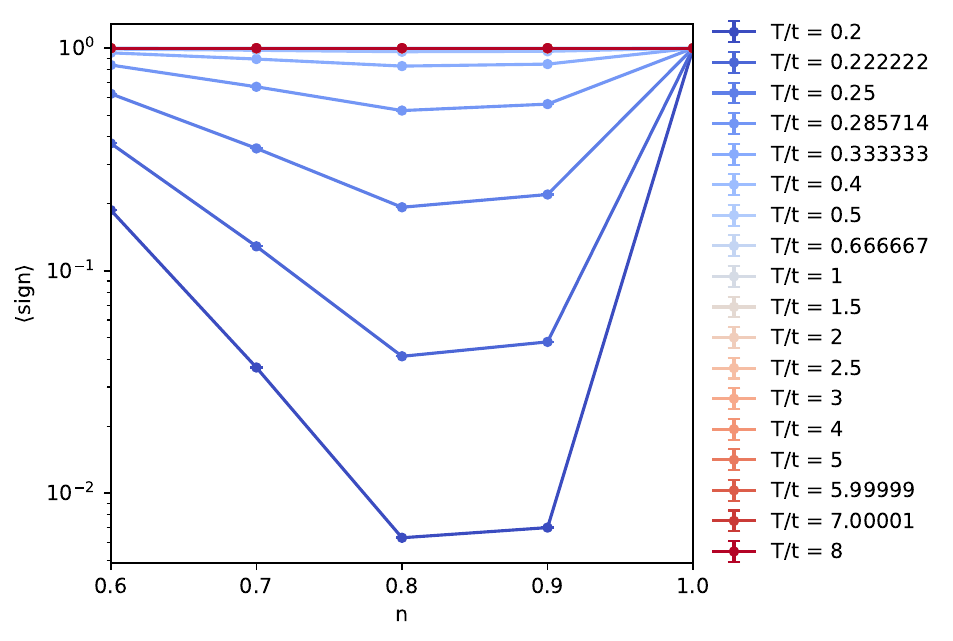}
    \caption{Average sign of the DQMC simulations of the repulsive Hubbard model at dopings of $\left\langle n \right\rangle = 0.6, 0.7, 0.8, 0.9$ and $1.0$, with $U/t=6$.}
    \label{fig:sign}
\end{figure}

Though the attractive Hubbard model in any lattice geometry is sign-problem-free, DQMC simulations of the repulsive Hubbard model exhibit a fermion sign problem upon doping. The average sign in our simulations for the repulsive Hubbard model is reported in \cref{fig:sign}. The measured average sign is exactly $1$ for $T\geq 1$.

\subsection{Finite-size convergence test}
\begin{figure}[!htbp]
    \centering
    \includegraphics[width=0.5\linewidth]{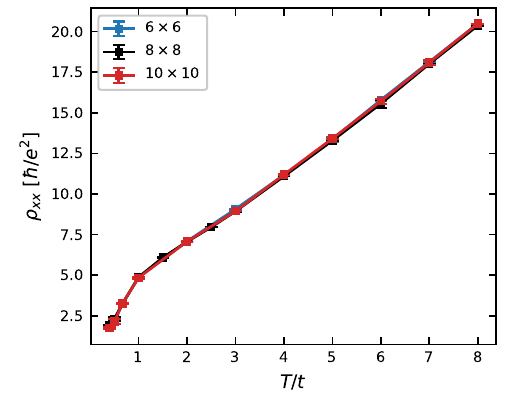}
    \caption{DC resistivity of the attractive Hubbard model at half-filling, with $U/t = -6$ and different lattice sizes ($6\times 6$, $8\times 8$, and $10\times 10$).}
    \label{fig:resistivity_finite_size}
\end{figure}

Comparisons of longitudinal DC resistivity of the attractive Hubbard model with different lattice size ($6\times 6$, $8\times 8$, and $10\times 10$) are shown in \cref{fig:resistivity_finite_size}. Due to the considerable computational expense of larger cluster simulations, we focus on the half-filling case with $\left\langle n\right\rangle=1$. Since the data of conductivity and resistivity in all cases are quantitatively similar, the data presented in the main text on an $8\times 8$ lattice are void of significant finite-size effects.


\subsection{Monte Carlo sweep convergence test}
\begin{figure}[!htbp]
    \centering
    \includegraphics[width=0.5\linewidth]{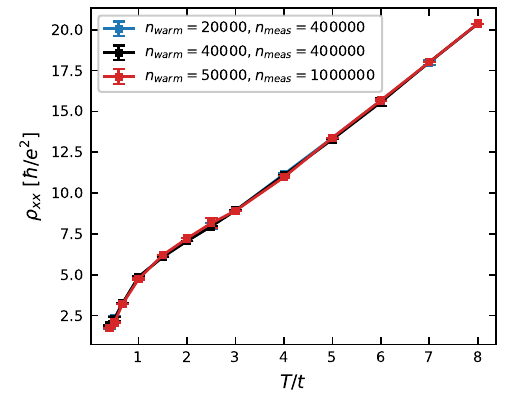}
    \caption{Convergence of the DC resistivity of the half-filled attractive Hubbard model with respect to the numbers of warm-up and measurement sweeps. The calculations are performed on an $8\times8$ lattice with $U/t=-6$.}
    \label{fig:resistivity_sweep_convergence}
\end{figure}

Comparison of DC resistivity of the attractive Hubbard model obtained with different number of warm-up and measurement sweeps are shown in \cref{fig:resistivity_sweep_convergence}. Due to the considerable computational expense of larger number of sweeps, we focus on the half-filling case with $\left\langle n\right\rangle=1$. Since the data of resistivity in all case are quantitatively similar, the data presented in the main text are obtained with $4\times 10^{4}$ warm-up sweeps and $4\times 10^{5}$ measurement sweeps, with which the measured properties converge within the margin of error.

\subsection{Ground states of attractive and repulsive Hubbard models}
\begin{figure}[!htbp]
    \centering
    \includegraphics[width=\linewidth]{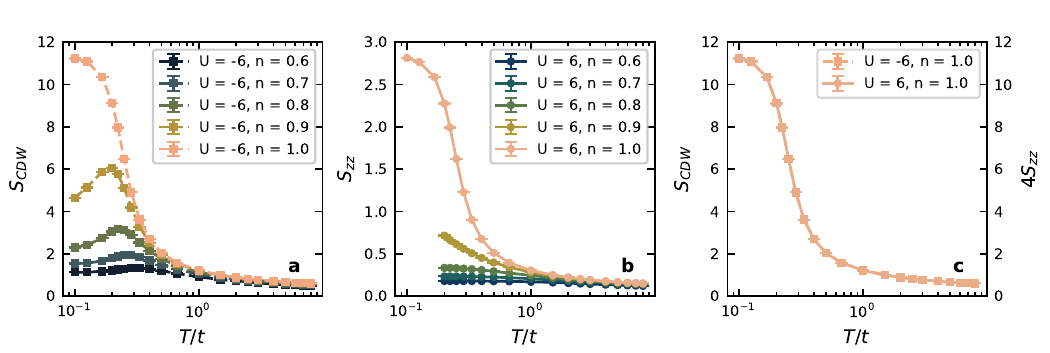}
    \caption{Charge and spin structure factor for $\left|U\right|/t = 6$, $\left\langle n \right\rangle = 0.6, 0.7, 0.8, 0.9$ and $1.0$. \textbf{a} Charge structure factor at $\left(\pi,\pi\right)$ for $U/t = -6$. \textbf{b} Spin structure factor at $\left(\pi,\pi\right)$ for $U/t = 6$. \textbf{c} Comparison of the charge structure factor at $\left(\pi,\pi\right)$ for $U/t = -6$ and the scaled spin structure factor at $\left(\pi,\pi\right)$ for $U/t = 6$.}
    \label{fig:structure_factor}
\end{figure}

To illustrate the differences in the ground-state properties of the attractive and repulsive Hubbard model, we examine their equal-time structure factors for spin and charge. \cref{fig:structure_factor} shows the checkerboard charge-density-wave (CDW) structure factor
\begin{equation}
    S_{\mathrm{CDW}}\left(\pi ,\pi \right)=\sum_{\mathbf{r}} \left(-1\right)^{r_{x}+r_{y}}\left( \left< n_{\mathbf{0}}n_{\mathbf{r}} \right> -\left< n \right>^{2} \right)
\end{equation}
of the attractive Hubbard model with and the $z$-component of the antiferromagnetic spin structure factor at $\mathbf{Q}=\left(\pi ,\pi \right)$ 
\begin{equation}
    S_{zz}\left( \pi ,\pi \right) =\sum_{\mathbf{r}} \left( -1 \right)^{r_{x}+r_{y}} \left( \left< S_{\mathbf{0}}^{z}S_{\mathbf{r}}^{z} \right> -\left< S^{z} \right>^{2} \right)
\end{equation}
of the repulsive Hubbard model over a wide range of doping. As the temperature decreases, both $S_{\mathrm{CDW}}\left(\pi,\pi\right)$ in the attractive Hubbard model and $S_{zz}\left( \pi ,\pi \right)$ in the repulsive Hubbard model increase markedly at half-filling, indicating the enhancement of checkerboard charge correlations and antiferromagnetic spin correlations, respectively. Upon hole doping, both structure factors are suppressed relative to the half-filled values. At half-filling, $S_{\mathrm{CDW}}(\pi,\pi)$ in the attractive Hubbard model overlaps, within numerical precision, with $4S_{zz}(\pi,\pi)$ in the repulsive model, as shown in \cref{fig:structure_factor}\textbf{c}, where the factor of $4$ accounts for the different operator normalizations. This agreement is consistent with the exact particle-hole mapping between charge (pseudospin) correlations in the attractive model and spin correlations in the repulsive model.

\subsection{Charge transport}
\begin{figure*}[!htbp]
    \centering
    \includegraphics[width=\linewidth]{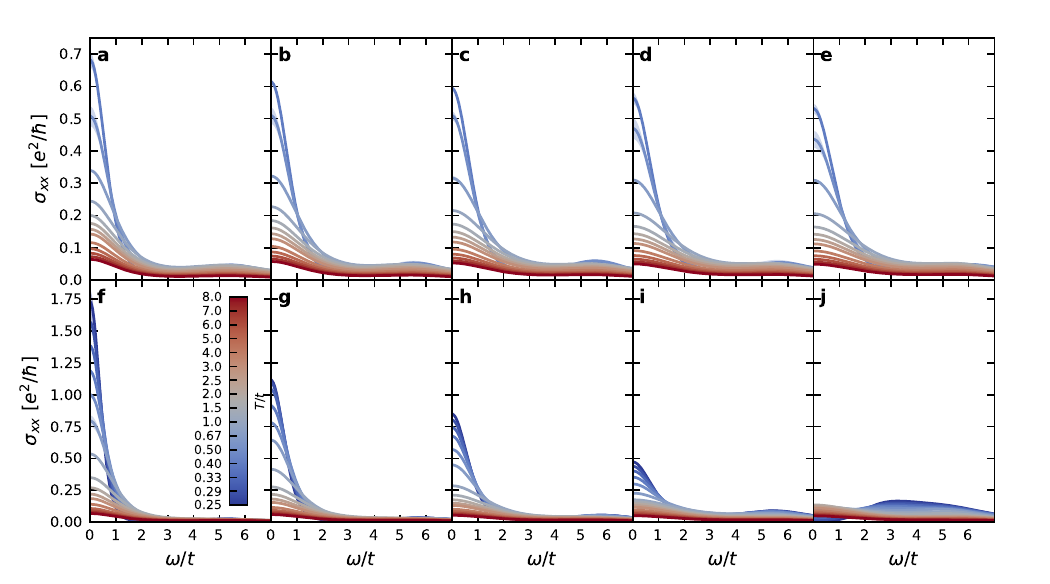}
    \caption{Optical conductivity of the attractive (\textbf{a} - \textbf{e}) and repulsive (\textbf{f} - \textbf{j}) Hubbard model under different temperatures $T = 1/\beta$, obtained through DQMC and MaxEnt analytical continuation, with $\left|U\right|/t = 6$.}
    \label{fig:conductivity}
\end{figure*}

\begin{figure}[!htbp]
    \centering
    \includegraphics[width=0.5\linewidth]{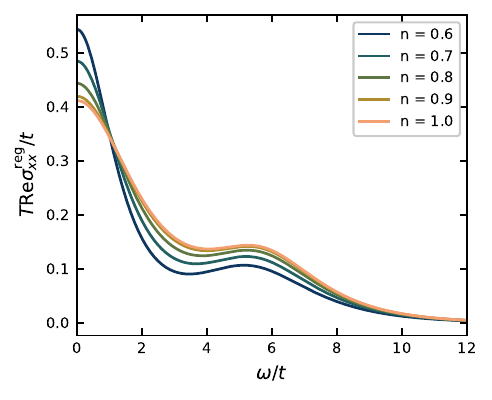}
    \caption{Scaled optical conductivity $T\sigma_{xx}$ as a function of frequency for $\left|U\right|/t=6$, $\left\langle n \right\rangle = 0.6, 0.7, 0.8, 0.9$ and $1.0$. The results are reconstructed from infinite-temperature expansion moments using Padé approximation and hyperbolic-function analytic continuation.}
    \label{fig:high_T_Tsigma}
\end{figure}

\begin{figure*}[!htbp]
    \centering
    \includegraphics[width=\linewidth]{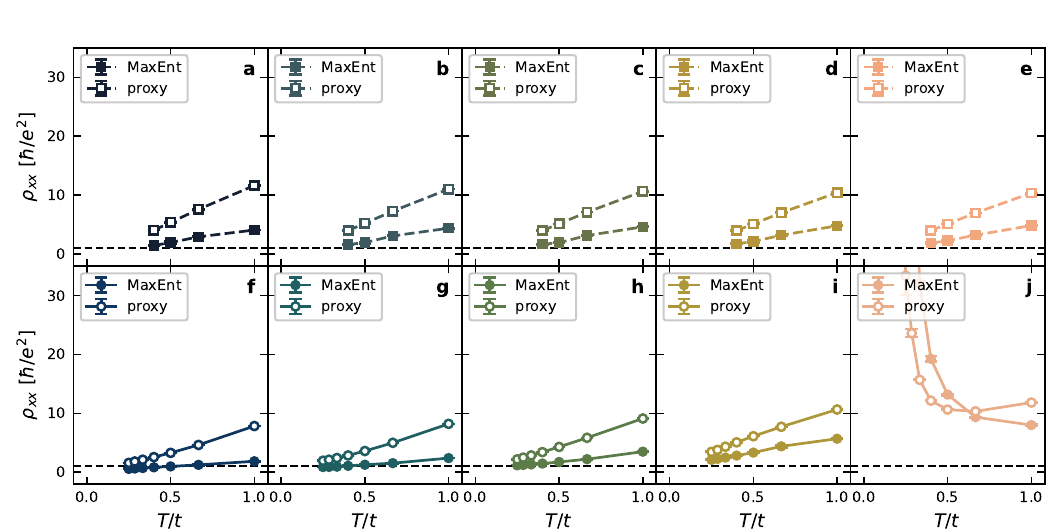}
    \caption{Comparison of DC resistivities of the attractive and repulsive Hubbard model under different temperatures $T = 1/\beta$, calculated with MaxEnt analytical continuation and the resistivity proxy $\rho_{\text{proxy},1}=\pi T^{2}/\Lambda\left(\beta/2\right)$, with $\left|U\right|/t = 6$. The Ioffe-Regel limit ($\rho_{xx} = \hbar/e^{2}$) is shown in black dashed line.}
    \label{fig:resistivity_proxy1}
\end{figure*}

\begin{figure*}[!htbp]
    \centering
    \includegraphics[width=\linewidth]{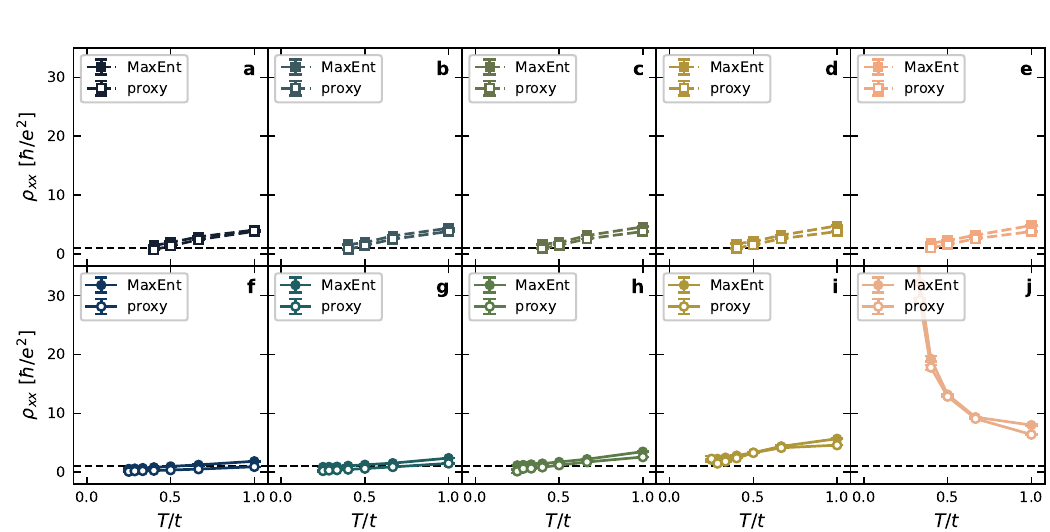}
    \caption{Comparison of DC resistivities of the attractive and repulsive Hubbard model under different temperatures $T = 1/\beta$, calculated with MaxEnt analytical continuation and the resistivity proxy $\rho_{\text{proxy},2}=\Lambda^{\prime\prime}\left(\beta/2\right)/\left(2\pi\Lambda\left(\beta/2\right)^{2}\right)$, with $\left|U\right|/t = 6$. The Ioffe-Regel limit ($\rho_{xx} = \hbar/e^{2}$) is shown in black dashed line.}
    \label{fig:resistivity_proxy2}
\end{figure*}

\cref{fig:conductivity} illustrates the frequency dependence of the longitudinal conductivity $\sigma_{xx}\left(\omega\right)$ under different temperature of the attractive and repulsive Hubbard models. \cref{fig:high_T_Tsigma} shows the scaled regular conductivity $T\mathrm{Re}\ \sigma_{xx}^{\mathrm{reg}}$ obtained from the infinite-temperature series expansion for all tested dopings. Two proxies of the DC resistivity, which are respectively
\begin{equation}
    \rho_{\text{proxy},1}=\frac{\pi T^{2}}{\Lambda\left(\beta/2\right)},
\end{equation}
and
\begin{equation}
    \rho_{\text{proxy},2}=\frac{\Lambda^{\prime\prime}\left(\beta/2\right)}{2\pi\Lambda\left(\beta/2\right)^{2}},
\end{equation}
are shown together with the DC resistivity $\rho_{\mathrm{ME}}$ obtained with the MaxEnt analytical continuation in \cref{fig:resistivity_proxy1} and \cref{fig:resistivity_proxy2}. The DC resistivity $\rho_{\mathrm{ME}}$ is consistent qualitatively with both of the proxies. Compared with $\rho_{\text{proxy},1}$, $\rho_{\text{proxy},2}$ normally provides a more robust estimate of resistivity

\bibliography{main}